

\def\ctb#1{\hfill #1 \hfill}
\input lecproc.cmm
\input psfig
\contribution{Summary of $\pi$--$\pi$ Scattering Experiments}

\author {Dinko Po\v{c}ani\'c}
\address {Department of Physics, University of Virginia \newline
Charlottesville, VA 22901-2458, USA 
         }

\abstract {
The $\pi\pi$ scattering amplitude at threshold is fully determined by the
chiral symmetry breaking part of the strong interaction, and, thus, directly
constrains the form of the low energy effective lagrangians.  Current status
of the study of the low energy $\pi\pi$ interaction is discussed,
particularly the recent results on reactions $\pi N \rightarrow\pi\pi N$ near
threshold.  Present levels of experimental uncertainties and limitations
inherent to the available analysis methods leave ample room for improvements
in the determination of the s-wave $\pi\pi$ scattering lengths.  Experimental
improvements are expected from new measurements of $K_{e4}$ decays and from
attempts to study $\pi^+\pi^-$ atoms, while further theoretical work is
required in order to make full use of the extensive near-threshold $\pi N
\rightarrow \pi\pi N$ data that has recently become available. }

\titlea{1}{Introduction}

Effects of chiral symmetry breaking in low energy interactions of pions, and
light hadrons in general, have been studied for over thirty years.  Strong
interactions break chiral symmetry both ``spontaneously'' and explicitly.
Spontaneous breaking of chiral symmetry is well understood and has led to new
concepts and methods that have transcended the domain of intermediate energy
physics.  On the other hand, the precise mechanism of explicit chiral
symmetry breaking (ChSB) remained largely unresolved until the establishment
of QCD as the correct theory of strong interactions.  The study of explicit
ChSB, however, remains relevant even today, owing to the failure of the full
QCD to describe strong interaction phenomena at energies below a few GeV.  At
these energies, QCD becomes nonperturbative and intractable in practice by
available calculational methods.  In order to overcome this problem, a broad
theoretical effort is under way to develop phenomenological lagrangian models
based solely on the symmetry properties of the full QCD, as suggested by
Weinberg [1].  Chiral symmetry plays a particularly important role at low
energies, since it is violated only slightly in the SU(2) realization, and is
essential for the understanding of the lightest hadrons.  For this reason,
chiral symmetry provides either the basic framework or the essential
constraints for all modern effective lagrangian models at low energies, such
as the chiral perturbation theory (ChPT) [2], and the various realizations of
the Nambu--Jona-Lasinio model [3].

It has been long established [4] that low energy $\pi\pi$ scattering provides
a particularly sensitive tool in studying the explicit breaking of chiral
symmetry, since $a(\pi\pi)$, the $\pi\pi$ scattering lengths, vanish exactly
in the chiral limit.  To the extent that they differ from zero, $a(\pi\pi)$
provide a direct measure of the symmetry breaking term in the pion sector.
Stated in more contemporary language, detailed knowledge of the $\pi\pi$
interaction scattering lengths and phase shifts provides much needed input in
fixing the parameters of ChPT and other effective models.


\titlea{2} {Experimental Determination of $\pi\pi$ Scattering Lengths}

Experimental evaluation of $\pi\pi$ scattering observables is difficult,
primarily because free pion targets are not available.  Scattering lengths
are especially hard to determine since they require measurements close to the
$\pi\pi$ threshold, where the available phase space strongly limits
measurement rates.  Over the years several reactions have been studied or
proposed as a means to obtain near-threshold $\pi\pi$ phase shifts, such as
$\pi N\rightarrow \pi\pi N$, $K_{e4}$ decays, $\pi^+\pi^-$ atoms,
$e^+e^-\rightarrow \pi\pi$, etc.  In practice, only the first two reactions
have so far proven useful in studying threshold $\pi\pi$ scattering.  The
three main methods are discussed below in the order of decreasing
reliability.

\titleb{2.1} {Analysis of $K\rightarrow\pi^+\pi^- e^+\nu$ decays}

The $K^+\rightarrow\pi^+\pi^- e^+\nu$ decay (called $K_{e4}$) is in several
respects the most suitable process for the study of near-threshold $\pi\pi$
interactions.  The interaction takes place between two real pions on the mass
shell, the only hadrons in the final state.  The dipion invariant mass
distribution in $K_{e4}$ decay peaks close to the $\pi\pi$ threshold, and
$l=I=0$ and $l=I=1$ are the only dipion quantum states contributing to the
process.  These factors, and the well understood $V-A$ nature of the weak
decay, favor the $K_{e4}$ process among all others in terms of theoretical
uncertainties.  On the other hand, measurements are impeded by the low
branching ratio of the decay, $\sim 3.9\times 10^{-5}$.

The most recent $K_{e4}$ measurement was made by a Geneva--Saclay
collaboration in the mid-1970's [5].  Using a detector system consisting of
pion \v{C}erenkov counters, wire chambers, a bending magnet and plastic
scintillator hodoscopes around a 4 m decay zone, Rosselet and coworkers
detected some 30,000 $K_{e4}$ decays.  Analysis of this low-background,
high-statistics data illustrates well the difficulties encountered in
extracting experimental $\pi\pi$ scattering lengths.  Figure~1. summarizes
the $\pi\pi$ phase shift information below 400 MeV derived from all $K_{e4}$
measurements to date.  The curves in Fig.~1 correspond to three different
values of $a_0^0(\pi\pi)$, and illustrate the relative insensitivity of the
data at the present level of experimental accuracy.

\midinsert
\vbox{\centerline{\psfig{figure=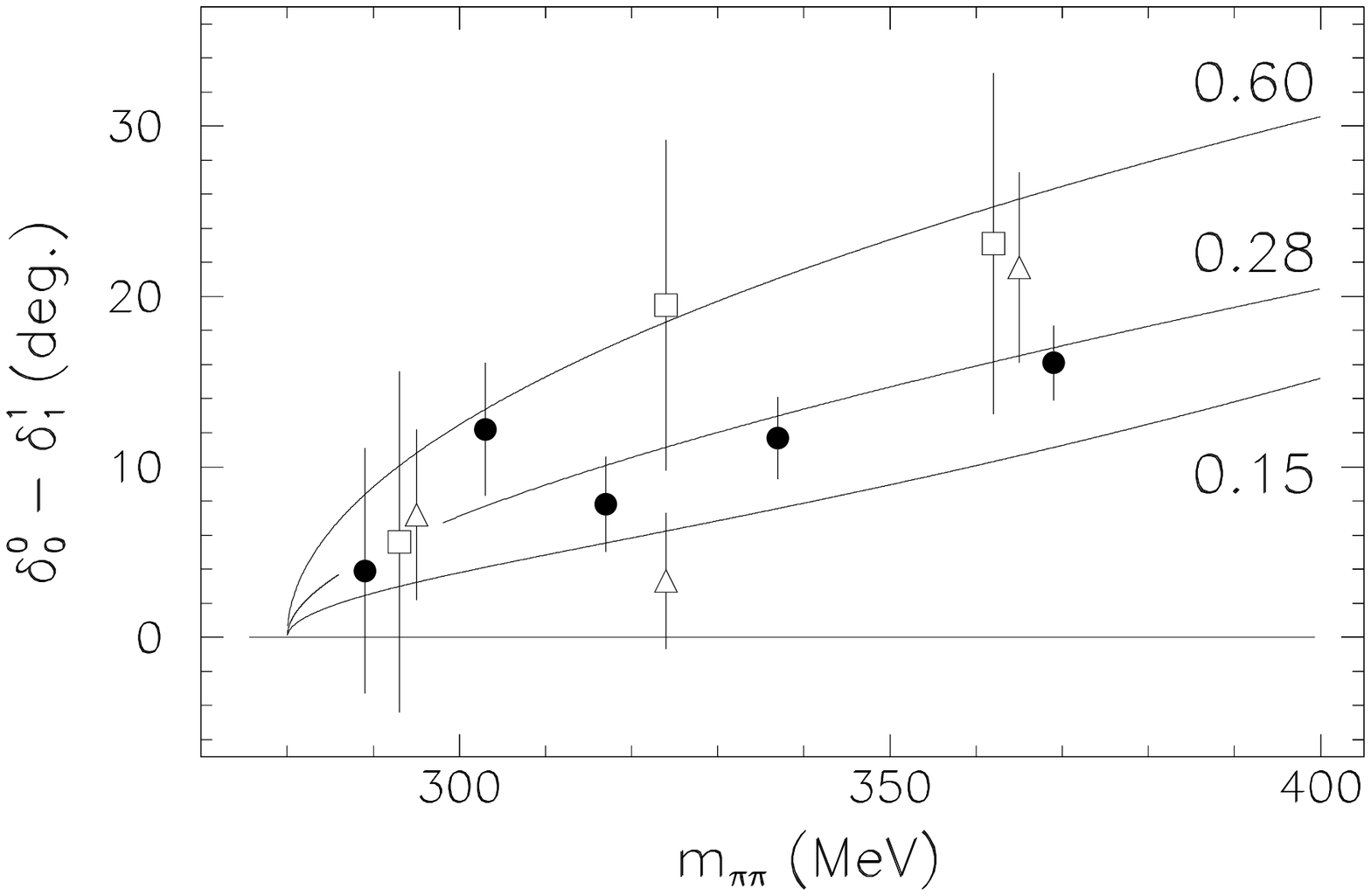,height=80mm}} }
\figure{1} {$\pi\pi$ phase shift information extracted from studies of
$K_{e4}$ decays, after Rosselet {\it et al.} [5].  Phase-shift difference
$\delta^0_0 - \delta^1_1$ is plotted against $m_{\pi\pi}$, the dipion
invariant mass.  Full circles represent results of Rosselet {\it et al.} [5],
while open squares and triangles represent the results of Zylberstejn [41]
and Beier {\it et al.} [42], respectively.  The three curves correspond to
phase shift solutions assuming three different values of $a_0^0$, as noted.}
\endinsert

By itself, the Geneva--Saclay experiment determines $a_0^0$ with a $\sim$35\%
uncertainty, and constrains $b$:
$$ a_0^0 = 0.31 \pm 0.11\ \mu^{-1} \quad ,\hskip 5mm b = b_0^0 - a_1^1 = 0.11
\pm 0.16 \ \mu^{-1} \quad ,$$
where $b_0^0$ is the s-wave $I=0$ slope parameter defined in the usual way by
$$
{{\rm Re} A_l^I \over q^{2l}} = a_l^I + b_l^Iq^2 +O(q^4) \quad , {\hskip 5mm}
{\rm with} {\hskip 5mm} q={1\over 2}\sqrt{s-4\mu^2}\quad , \eqno (1)
$$
where $A_l^I$ is the $\pi\pi$ partial wave amplitude, $s$ is the center
of mass energy of the two pions, and $\mu\equiv m_\pi$ is the pion mass.

$\pi\pi$ scattering phase shifts are, however, further constrained by
unitarity, analyticity, crossing and Bose symmetry, extensively studied by
Roy [6], and Basdevant {\it et al.} [7,8].  These constraints are expressed
in a set of dispersion relations known as the ``Roy equations'', which have
been evaluated on the basis of existing peripheral $\pi N \rightarrow\pi\pi
N$ data (see Sect. 2.2).  By applying the Roy equation constraints of
Basdevant, {\it et al.}  [8], and thus combining the $K_{e4}$ and
peripheral pion production results, more accurate values for $a_0^0$ and
$b_0^0$ were obtained [5], as well as values of scattering lengths and slope
parameters for ($l=0$, $I=2$) and ($l=1$, $I=1$) [9].

The present experimental accuracy of the $K_{e4}$ measurement of the $\pi\pi$
phase shifts clearly needs to be improved.  It is also evident that an
independent accurate experimental determination of $a_0^2$, the $I=2$ s-wave
scattering length, is called for, since this quantity is not directly
constrained by $K_{e4}$ measurements.

\titleb{2.2} {Peripheral $\pi N \rightarrow \pi \pi N$ Reactions: the
Chew--Low Method}

Particle production in peripheral collisions can be used to extract
information on the scattering of two of the particles in the final state, as
shown by Chew and Low in 1959 [10].  Applied to the $\pi N \rightarrow \pi
\pi N$ reaction, the well-known Chew--Low formula,
$$
\sigma_{\pi\pi}(m_{\pi\pi}) = \lim_{t\rightarrow\mu^2}\ \left[
{\partial^2\sigma_{\pi\pi N} \over \partial t\partial m_{\pi\pi} } \cdot {\pi
\over \alpha f_\pi^2} \cdot {p^2(t-\mu^2)^2 \over tm_{\pi\pi}k} \right]
\quad , \eqno (2)
$$
relates $\sigma_{\pi\pi}(m_{\pi\pi})$, the cross section for pion-pion
scattering, to double differential $\pi N \rightarrow \pi\pi N$ cross section
and kinematical factors: $p$, momentum of the incident pion, $m_{\pi\pi}$,
the dipion invariant mass, $t$, the Mandelstam square of the 4-momentum
transfer to the nucleon, $k=(m_{\pi\pi}^2/4-\mu^2)^{1/2}$, momentum of the
secondary pion in the rest frame of the dipion, $f_\pi \approx 93$ MeV, the
pion decay constant, and $\alpha=1$ or 2, a statistical factor involving the
pion and nucleon charge states.  The method relies on an accurate
extrapolation of the double differential cross section to the pion pole in
order to isolate the one pion exchange (OPE) pole term contribution.  Since
the exchanged pion is off-shell in the physical region ($t<0$), this method
requires measurements under conditions which maximize the OPE contribution
and minimize all background contributions.  Thus, suitable measurements
require peripheral pion production at values of $t$ as close to zero as
possible, which become available at incident momenta typically above $\sim$ 3
GeV/c.

The Chew--Low method has been refined considerably over time, particularly by
Baton and coworkers [11], whose approach enables extraction of $\pi\pi$ phase
shifts through appropriate treatment of the angular dependence of the $\pi
N\rightarrow \pi\pi N$ exclusive cross sections.  Crossing, Bose and isospin
symmetries, analyticity and unitarity, provide dispersion relation
constraints on the $\pi\pi$ phase shifts, the ``Roy equations'' [6-8].  These
constraints are particularly useful in evaluating $\pi\pi$ scattering lengths
because available phase space restricts severely the statistics of
peripheral $\pi N\rightarrow\pi\pi N$ measurements below $m_{\pi\pi} \approx
500$ MeV, while accurate data are available at higher $\pi\pi$ energies.

The data base for these analyses has essentially not changed since the early
1970's, and is dominated by two experiments, performed by the Berkeley [12]
and CERN-Munich [13] groups.  The latter of the two measurements has much
higher statistics (300 k events compared to 32 k in the Berkeley experiment).
These data were subsequently analyzed by numerous authors, too many to review
here; ultimately, the peripheral pion production results were combined with
the Geneva--Saclay $K_{e4}$ data in a comprehensive dispersion-relation
analysis [9], as discussed in Sect.~2.1.

There have been independent Chew--Low type analyses since the 1970's; the
last published one, performed by the Kurchatov Institute group in 1982,
was based on a set of some 35,000 events recorded in bubble chambers [14].
The same group has recently updated their analysis [15].

\titleb{2.3} {$\pi N \rightarrow \pi \pi N$ Reactions near Threshold}

Early on, Weinberg pointed out that the OPE graph dominates the $\pi N
\rightarrow \pi \pi N$ reaction at threshold [4].  Subsequently, Olsson and
Turner constructed a soft-pion lagrangian containing only the OPE and contact
terms at threshold [16], and derived a straightforward relation between the
$\pi\pi$ and $\pi N\rightarrow \pi \pi N$ threshold amplitudes with only one
parameter, $\xi$, the chiral symmetry breaking parameter.  Thus, in
Olsson--Turner's model, it is sufficient to measure total $\pi N \rightarrow
\pi\pi N$ cross sections, from which quasi-amplitudes can be calculated and
extrapolated to threshold.  In spite of recent strong criticism for
incompatibility with QCD and oversimplified dynamical assumptions, Olsson and
Turner's work to date provides the sole direct relation between $\pi
N\rightarrow\pi\pi N$ observables and the $\pi\pi$ lagrangian.  In this way,
soft-pion theory has provided the main inspiration for the near-threshold
$\pi N\rightarrow\pi\pi N$ measurements and, in spite of its shortcomings, is
still being used by experimentalists to relate the results from different
reaction channels in a systematic way.

Unlike the methods described in Sects. 2.1 and 2.2, the last ten years have
witnessed a great deal of experimental activity on exclusive and inclusive
near-threshold $\pi N\rightarrow\pi\pi N$ measurements.  As in Chew--Low
peripheral pion production, there are 5 charge channels accessible to
measurement,
$$
\pi^-p\rightarrow\pi^-\pi^+n \quad , \hskip 10mm
\pi^-p\rightarrow\pi^0\pi^0n \quad , \hskip 10mm
\pi^-p\rightarrow\pi^-\pi^0p \quad , $$
$$
\pi^+p\rightarrow\pi^+\pi^0p \quad , \hskip 10mm {\rm and} \hskip 10mm
\pi^+p\rightarrow\pi^+\pi^+n \quad .
$$
For brevity, we label them with their final state charges as ($-$+n), ...,
(++n), respectively.  Recent experiments reporting total $\pi
N\rightarrow\pi\pi N$ cross sections are summarized below, while data
available before 1984 is reviewed in Ref.~[17].

The OMICRON group at CERN has measured cross sections in the (+$-$n),
($-$0p), (++n) and (+0p) channels [18].  They detected coincident
pairs of charged particles in a two-sided magnetic spectrometer, restricted
to in-plane kinematics.  (Limitations inherent in the extraction of total
cross sections from data restricted to in-plane kinematics were recently
discussed by the Erlangen group [19].)  The thin gas target,
limited magnetic spectrometer acceptance, and background subtraction, result
in large error bars for some of the low-energy OMICRON data points,
particularly in the (++n) channel.

At TRIUMF, Sevior and coworkers measured inclusive cross sections for the
reaction (++n) using a novel technique involving an active plastic
scintillator target combined with neutron detection [20].  Total cross
sections were evaluated assuming s-wave dominance due to the proximity of the
threshold (5 MeV below their lowest energy measurement).  It is significant
that the TRIUMF results disagree with the (++n) OMICRON data.

J. Lowe and coworkers measured the (00n) channel at Brookhaven using the
Crystal Box detector [21].  Due to the large solid-angle coverage of
the detector, this was a kinematically complete measurement of 4 photons
following the decays of the two $\pi^0$'s in the final state.  As all
particles in the final state are neutral, the lowest point was also about 5
MeV above the threshold.

Finally, a Virginia-Stanford-LAMPF team studied the (+0p) channel using the
LAMPF $\pi^0$ spectrometer and an array of plastic scintillation telescopes
for $\pi^+$ and $p$ detection [22,23].  Three classes of exclusive events
were recorded simultaneously: $\pi^+\pi^0$ and $\pi^0p$ double coincidences,
and $\pi^+\pi^0p$ triple coincidences.  Since the acceptance of the apparatus
and the backgrounds were significantly different for the three classes of
events, this experiment had a strong built-in consistency check.

Published total cross sections for all five channels are summarized in
Fig.~2., shown in the form of quasi-amplitudes, $|a(\pi\pi N)|$, extracted
from the total cross sections following the prescription of Olsson and
Turner [16]
$$
\sigma(\pi N\rightarrow\pi\pi N) = |a(\pi\pi N)|^2\cdot\alpha\cdot
p_\pi^2\times{\rm phase\ \ space} \quad , \eqno (3)
$$
where $p_\pi$ is the c.m. incident pion momentum, and the statistical factor
$\alpha$ = 1/2 for the (++n) and (00n) channels, and $\alpha$ = 1 for the
other three channels.
\midinsert
\vbox{
\centerline{\psfig{figure=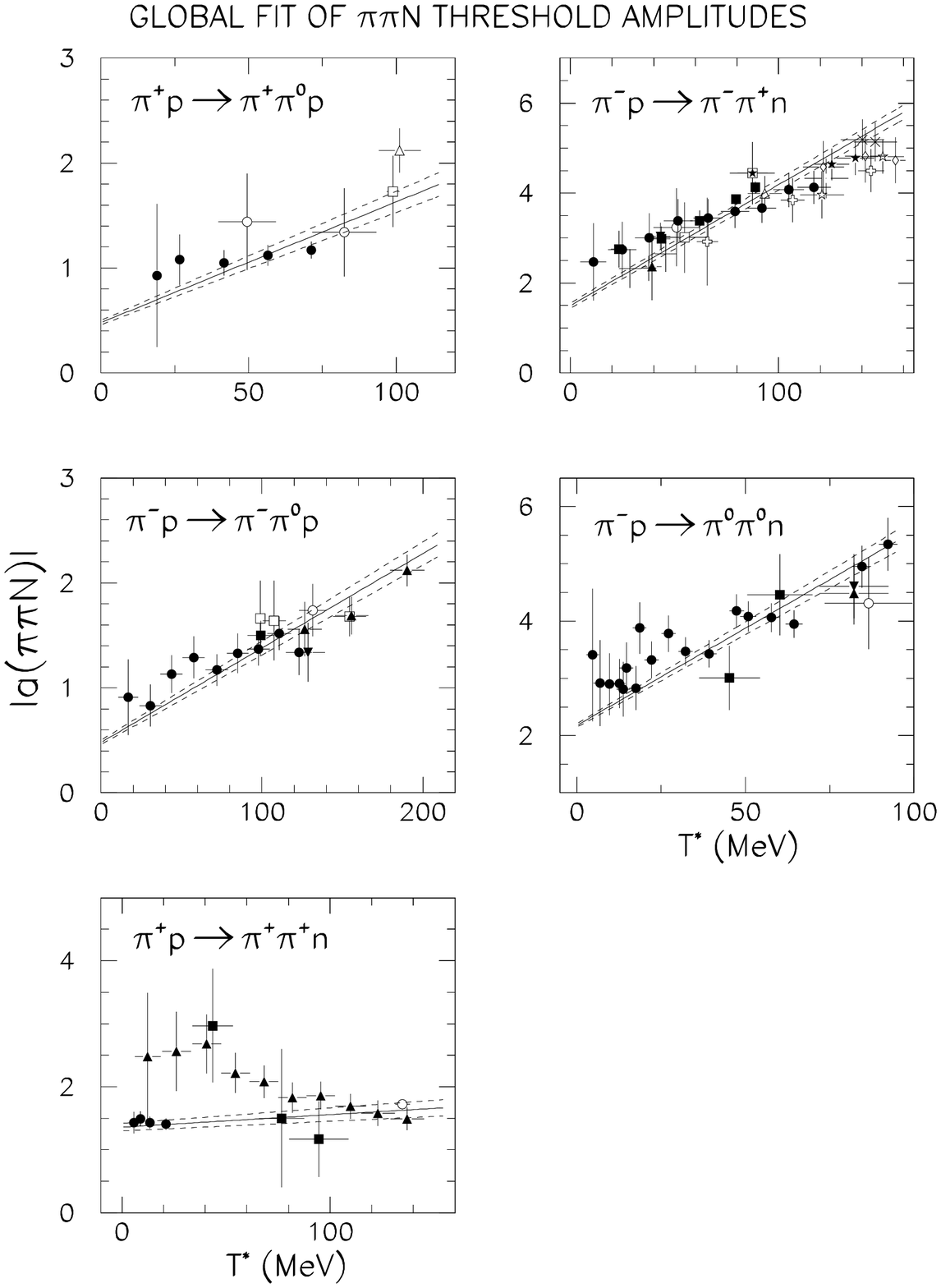,height=180mm}}
     }
\vskip -50mm \hskip 58mm \vbox{\hsize=57mm
\figure{2} {Compilation of low energy data on $\pi N\rightarrow\pi\pi N$
quasi-ampli\-tudes plotted against $T^*$, the c.m. energy above the channel
threshold.  Solid lines: results of a global fit performed by the Virginia
group using the soft-pion model constraints of Olsson and Turner.  Dashed
lines correspond to fits with $\chi^2_{\rm min}+1$.}
\vskip 5mm                   }
\endinsert

Apart from the pronounced disagreement in the (++n) channel between the
TRIUMF [20] and OMICRON [18] data, represented in Fig.~2. by full circles and
full triangles, respectively, the entire body of data appears globally
consistent within the quoted uncertainties.  The straight lines drawn through
the data in Fig.~2 are the result of a constrained soft-pion analysis
following Olsson and Turner performed by the Virginia group, with the result
of $\xi = -0.25\pm 0.10$ [22,23].  A similar global fit was performed
earlier by Burkhardt and Lowe using a slightly different fitting procedure,
yielding $\xi = -0.60 \pm 0.10$ [24].  In both analyses the quoted
uncertainty of $\xi$ is determined only by the experimental errors and the
statistical quality of the global fit.  We can interpret the spread between
the two values as due to the systematic uncertainties of the method, and take
the mean as a representative soft-pion analysis result.

The body of near-threshold $\pi N\rightarrow\pi\pi N$ data keeps growing.
Several experiments presently under way at TRIUMF and LAMPF are expected to
yield new results shortly, on both exclusive and inclusive cross sections in
the (+$-$n), (++n), (+0p) and ($-$0p) channels.  We also note the
high-statistics angular correlation measurements in the (+$-$n) channel
performed by the Erlangen group, who did not report total cross sections
[25].

Closer analysis of the exclusive cross sections is only now beginning, as
high-statistics data have not been available until recently, and the methods
of analysis are still being refined.  In a series of recent papers, the
Erlangen group has focused on the main graphs contributing to the continuum
$\pi N\rightarrow \pi\pi N$ amplitude [26].  On the other hand, the
St. Petersburg group [27] has derived the most general constraints on the
kinematical dependence of the background and the OPE amplitudes in the
physical region, as dictated by basic symmetries: crossing, Bose, isospin,
analyticity and unitarity.

\titlea{3} {Comparison of Experimental Results and Predictions}

In order to obtain a proper perspective on the existing data and the three
experimental methods discussed in the preceding sections, we review briefly
the various theoretical calculations of the s-wave $\pi\pi$ scattering
lengths, in the order in which they appeared.

Weinberg's soft-pion model of chiral symmetry breaking relied on current
algebra and PCAC [4,28].  Weinberg required of his lagrangian that
$\partial^\mu A_\mu$, the divergence of the axial-vector current, form a
chiral quadruplet with the pion field.  Consequently, the $\pi\pi$ part of
the lagrangian assumes the form:
$$
{\cal L}_{\pi\pi} = -{1 \over {4f_\pi^2}}\cdot [\vec{\phi}^2(\partial_\mu
\vec{\phi})^2 - {1\over 2}\mu^2(\vec{\phi}^2)^2] \quad , \eqno (4)
$$
where $\vec{\phi}$ is the pion field.  Weinberg also noted that the s-wave
scattering lengths $a_0^0$ and $a_0^2$ are constrained linearly:
$$
2a_0^0 - 5a_0^2 = 6\ {\mu\over {8\pi f_\pi^2}} \approx 0.56\ \mu^{-1} \quad .
\eqno (5)
$$
Predictions by Schwinger [29] and Chang and G\"ursey [30] differed from
Weinberg's in the form of the response of the pion field to chiral
transformations, resulting in different coefficients of the $\phi^4$ pion
mass term in (4).  Instead of the Weinberg's coefficient 1/2, Schwinger's
model suggests 1/4, and Chang and G\"ursey's 1/3, respectively.
Consequently, calculated values for the $\pi\pi$ s-wave scattering lengths
differ.

Since in soft-pion theory (5) constrains $a_0^0$ and $a_0^2$ linearly,
different models need only to fix the ratio $a_0^2/a_0^0$, i.e., only one
degree of freedom remains.  In this respect, Olsson and Turner's parameter
$\xi$ (see Sect.~2.3 and Ref.~[16]) determines the magnitude of the $\phi^4$
pion mass term in (4), and, consequently, the $\pi\pi$ scattering
lengths.  Thus, Weinberg's prediction is equivalent to setting $\xi=0$,
Schwinger's and Chang--G\"ursey's to $\xi$ = 1 and 2/3, respectively.
Although QCD has confirmed Weinberg's choice as the correct leading-order
term, we include the two latter results for historical completeness.

The current-algebra calculation of Weinberg was improved in 1982 by Jacob and
Scadron who introduced a correction due to the non-soft
$S^*\rightarrow\pi\pi$ isobar background [31].  At about the same time,
Gasser and Leutwyler calculated the $\pi\pi$ scattering amplitude to order
$p^4$ in ChPT [32], and gave scattering length predictions.  Also inspired by
QCD, Ivanov and Troitskaya used the model of dominance by quark loop
anomalies (QLAD) to obtain $\pi\pi$ scattering lengths [33].  On the other
hand, the J\"ulich group constructed a dynamical model of
pseudo\-scalar-pseudo\-sca\-lar meson scattering based on meson exchange, and
applied it to calculate a number of $\pi\pi$ and $K\pi$ scattering
observables at low and intermediate energies [34].  A somewhat complementary
approach to ChPT is the Nambu--Jona-Lasinio model [3].  Calculations of
$\pi\pi$ scattering lengths within the SU(2)$\times$SU(2) and
SU(3)$\times$SU(3) realizations of the NJL model are found in Refs.~[35] and
[36], respectively.

Most recently, Kuramashi and coworkers successfully applied quenched lattice
QCD on a 12$^3\times$20 lattice, and obtained $I=0$ and 2 $\pi\pi$ scattering
length values in the neighborhood of the older current-algebra
calculations [37].  Finally, Roberts {\it et al.} have recently
developed a model field theory, referred to as the global color-symmetry
model (GCM), in which the interaction between quarks is mediated by dressed
vector boson exchange [38].  The model incorporates dynamical chiral
symmetry breaking, asymptotic freedom, and quark confinement, and was applied
to calculate a number of low-energy observables in the pion sector of QCD.

Table~1. summarizes quantitatively the theoretical model
predictions and the experimental determinations of the s-wave $\pi\pi$
scattering lengths.  The same quantities are also displayed in Fig.~3.

\topinsert
\tabcap{1} {The s-wave $\pi\pi$ scattering lengths: compilation of theoretical
model predictions and of experimental analysis results.  The fourth column
lists the chiral symmetry breaking ``offset'' $2a_0^0-5a_0^2$ defined in
Eq.~(5).  Uncertainties were not entered for theoretical predictions, as they
are normally not quoted by authors.  For the few calculations that did
estimate uncertainties, they ranged from $\sim$5 to $\sim$10\%.  All values
are listed in units of inverse pion mass, $\mu^{-1}$. }

\line{\hrulefill}
\settabs \+ \quad Chew--Low PSA & \quad $0.248\pm 0.038$ \quad & $-0.047\pm
0.006$ \quad & \quad $0.68\pm 0.21$ \quad & [12,34] & \quad \cr
\+ \quad Model/Method & \ctb{$a_0^0\ (\mu^{-1})$} & \ctb{$a_0^2\ (\mu^{-1})$}
& \ctb{$2a_0^0 - 5a_0^2$} & \ctb{Ref.} & \cr
\line{\hrulefill}
\centerline{\it (a) Theoretical model predictions }
\+ \quad Weinberg       & \ctb{0.16} & \ctb{-0.046} & \ctb{0.56} &
         \ctb{[4]} & \cr
\+ \quad Schwinger      & \ctb{0.10} & \ctb{-0.069} & \ctb{0.56} &
         \ctb{[29]} & \cr
\+ \quad Chang-G\"ursey & \ctb{0.12} & \ctb{-0.062} & \ctb{0.56} &
         \ctb{[30]} & \cr
\+ \quad Jacob-Scadron  & \ctb{0.20} & \ctb{-0.029} & \ctb{0.56} &
         \ctb{[31]} & \cr
\+ \quad ChPT           & \ctb{0.20} & \ctb{-0.042} & \ctb{0.61} &
         \ctb{[32]} & \cr
\+ \quad QLAD           & \ctb{0.21} & \ctb{-0.060} & \ctb{0.72} &
         \ctb{[33]} & \cr
\+ \quad Meson exch'ge  & \ctb{0.31} & \ctb{-0.027} & \ctb{0.76} &
         \ctb{[34]} & \cr
\+ \quad NJL -- SU(2)   & \ctb{0.22} & \ctb{-0.074} & \ctb{0.81} &
         \ctb{[35]} & \cr
\+ \quad NJL -- SU(3)   & \ctb{0.26} & \ctb{-0.062} & \ctb{0.83} &
         \ctb{[36]} & \cr
\+ \quad Lattice QCD    & \ctb{0.22} & \ctb{-0.042} & \ctb{0.65} &
         \ctb{[37]} & \cr
\+ \quad GCM (fit 1)    & \ctb{0.16} & \ctb{-0.042} & \ctb{0.53} &
         \ctb{[38]} & \cr
\+ \quad GCM (calc.)    & \ctb{0.17} & \ctb{-0.048} & \ctb{0.58} &
         \ctb{[38]} & \cr
\smallskip
\centerline{\it (b) Results of experimental analyses}
\+ \quad $K_{e4}$ + Roy eq.& \ctb{$0.26\pm 0.05$} & \ctb{$-0.028\pm 0.012$} &
         \ctb{$0.66\pm 0.12$} & \ctb{[5,9]} & \cr
\+ \quad Chew--Low PSA   & \ctb{$0.24\pm 0.03$} & \ctb{$-0.04\pm 0.04$} &
         \ctb{$0.68\pm 0.21$} & \ctb{[14]} & \cr
\+ \quad Soft-pion/O-T   & \ctb{$0.188\pm 0.016$} & \ctb{$-0.037\pm 0.006$} &
         \ctb{$0.56\pm 0.04$} & \ctb{[24,22]} & \cr
\line{\hrulefill}
\endinsert

\midinsert
\vbox{\centerline{\psfig{figure=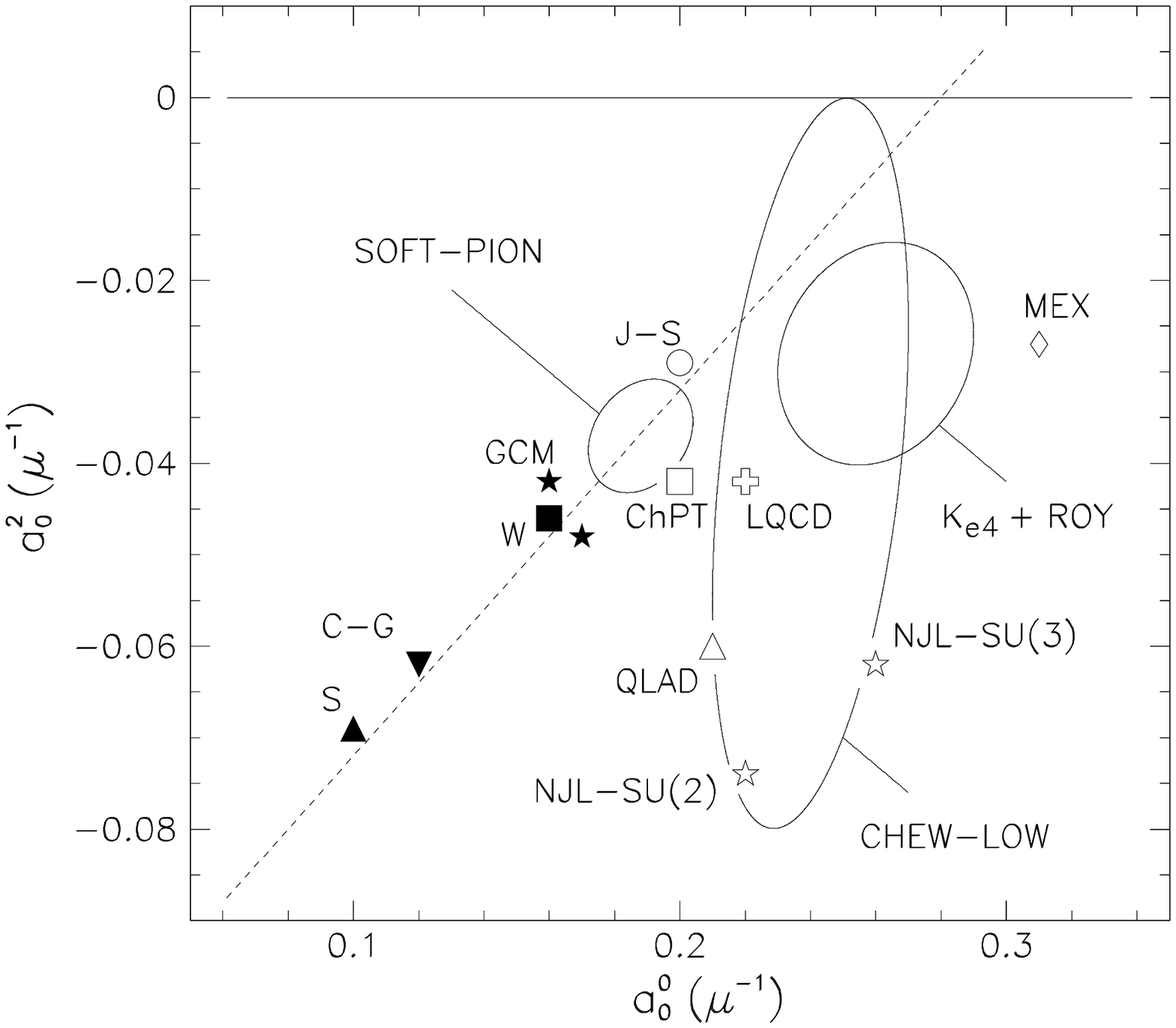,height=105mm}} }
\figure{3} {Summary of the $\pi\pi$ scattering length predictions (symbols)
and experimental results (contour limits).  Dashed line: Weinberg's
constraint given in Eq.~(5).  Numerical values of $\pi\pi$ scattering
lengths, experimental limits, and corresponding references are listed in
Table~1.  Model calculations: Weinberg (full square), Schwinger (filled
triangle), Chang and G\"ursey (filled inverted triangle), Jacob and Scadron
(open circle), Gasser and Leutwyler -- ChPT (open square), Ivanov and
Troitskaya -- QLAD (open triangle), Lohse {\it et al.} -- Meson Exchange
(open rhomb), Ruivo {\it et al.} and Bernard {\it et al.} -- NJL (open
stars), Kuramashi {\it et al.} -- quenched lattice gauge QCD (open cross),
Roberts {\it et al.} -- GCM (filled stars). }
\endinsert

Considerable scatter of predicted values of $a_0^0$ and $a_0^2$ is evident in
Fig.~3.  Even disregarding the 1960's calculations of Schwinger [29] and Chang
and G\"ursey [30] which were superseded by QCD, a significant range of
predicted values remains.  However, in view of the present experimental
uncertainties (see Fig.~3.), most authors claim that their results are
supported by the data.  This is not a satisfactory state of affairs.
Progress is required on two fronts in order to improve the present
uncertainties in the experimentally derived $\pi\pi$ scattering lengths.

First, more accurate data on $K_{e4}$ decays are needed, in order to reduce
the current error limits in the analysis.  However, $K_{e4}$ data alone will
not suffice because of their insensitivity to the $I=2$ $\pi\pi$ channel.
Measurements of the $\pi^+\pi^-$ atom proposed at several laboratories, if
feasible with reasonable statistics, could provide the much needed additional
theoretically unambiguous information.  The quantity to be measured is the
decay rate of the $\pi^+\pi^-$ ground state into the $\pi^0\pi^0$ channel,
which is proportional to $|a_0^0-a_0^2|^2$ [39].

Second, the mounting volume and accuracy of the exclusive and inclusive
near-threshold $\pi N\rightarrow \pi\pi N$ data contain, in principle,
valuable information regarding threshold $\pi\pi$ scattering.  Theoretical
interpretation of this data requires much improvement in order to make full
use of this information.  The work of Bernard, Kaiser and Mei{\ss}ner [40]
appears to be a promising step in that direction.
\medskip

This work has been supported by a grant from the United States National
Science Foundation.

\begrefchapter{References}
\refno {1.} S. Weinberg, Physica {\bf 96A} 327 (1979)

\refno {2.} J. Gasser and H. Leutwyler, Ann. Phys. (N.Y.) {\bf 158}
142 (1984); Nucl. Phys. {\bf B250} 465 (1985)

\refno {3.} Y. Nambu and G. Jona-Lasinio, Phys. Rev. {\bf 122} 345 (1961);
{\it ibid.} {\bf 124} 246 (1961); For a review of recent work based on the
NJL model see, e.g., S. P. Klevansky, Rev. Mod. Phys. {\bf 64} 649 (1992)

\refno {4.} S. Weinberg, Phys. Rev. Lett. {\bf 17} 616 (1966), {\it ibid.}
{\bf 18} 188 (1967)

\refno {5.}L. Rosselet, {\it et al.}, Phys. Rev. D {\bf 15} 574 (1977)

\refno {6.} S. M. Roy, Phys. Lett. {\bf35B} 353 (1971)

\refno {7.} J. L. Basdevant, J. C. Le Guillou and H. Navelet, Nuovo
Cim. {\bf 7A} 363 (1972)

\refno {8.} J. L. Basdevant, C. G. Froggatt and J. L. Peterson, Nucl.
Phys. {\bf B72} 413 (1974)

\refno {9.} M. M. Nagels,  {\it et al.}, Nucl. Phys. {\bf B147} 189
(1979)

\refno {10.} G. F. Chew and F. E. Low, Phys. Rev. {\bf 113} 1640 (1959)

\refno {11.} J. B. Baton, G. Laurens and J. Reignier, Phys. Lett. {\bf
33B} 525 (1970)

\refno {12.} S. D. Protopopescu {\it et al.}, Phys. Rev. D {\bf 7} 1279
(1973)

\refno {13.} G. Grayer {\it et al.}, Nucl. Phys. {\bf B75} 189 (1974)

\refno {14.} E. A. Alekseeva {\it et al.}, Zh. Eksp. Teor. Fiz. {\bf
82} 1007 (1982) [Sov. Phys. JETP {\bf 55} 591 (1982)]

\refno {15.} O. O. Patarakin and V. N. Tikhonov, Kurchatov Institute of
Atomic Energy preprint IAE-5629/2 (1993)

\refno {16.} M. G. Olsson and L. Turner, Phys. Rev. Lett. {\bf 20}
1127 (1968); Phys. Rev. {\bf 181} 2141 (1969), L. Turner, Ph. D. Thesis, Univ.
of Wisconsin, 1969 (unpublished)

\refno {17.} D. M. Manley, Phys. Rev. D {\bf 30} 536 (1984)

\refno {18.} G. Kernel,  {\it et al.},
Phys. Lett.  {\bf B216} 244 (1989); {\it ibid.} {\bf B225} 198 (1989); Z.
Phys. C {\bf 48} 201 (1990); {\it ibid.} {\bf 51} 377 (1991); in {\it
Particle Production Near Threshold} Nashville, 1990, edited by H. Nann and
E. J. Stephenson, (AIP, New York, 1991)

\refno {19.} H.-W. Ortner  {\it et al.} Phys . Rev. C {\bf 47} R447 (1993)

\refno {20.} M. E. Sevior,  {\it et al.} Phys. Rev. Lett. {\bf 66} 2569 (1991)

\refno {21.} J. Lowe,  {\it et al.} Phys. Rev. C {\bf 44} 956 (1991)

\refno {22.} D. Po\v{c}ani\'c, {\it et al.} Phys. Rev. Lett. {\bf 72} 1156
(1994)

\refno {23.} E. Frle\v{z}, Ph. D. Thesis, Univ. of Virginia, 1993 (Los
Alamos Report LA-12663-T, 1993)

\refno {24.} H. Burkhardt and J. Lowe, Phys. Rev. Lett. {\bf 67} 2622
(1991)

\refno {25.} H.-W. Ortner {\it et al.}, Phys. Rev. Lett. {\bf 64} 2759
(1990); R. M\"uller {\it et al.}, Phys. Rev. C {\bf 48} 981 (1993)

\refno {26.} O. J\"akel, H.-W. Ortner, M. Dillig and C. A. Z.
Vasconcellos, Nucl. Phys. {\bf A511} 733 (1990); O. J\"akel, M. Dillig and
C. A. Z. Vasconcellos, {\it ibid.} {\bf A541} 675 (1992); O. J\"akel and M.
Dillig, {\it ibid.} {\bf A561} 557 (1993)

\refno {27.} A. A. Bolokhov, V. V. Vereshchagin and S. G. Sherman, Nucl.
Phys. {\bf A530} 660 (1991)

\refno {28.} S. Weinberg, Phys. Rev. {\bf 166} 1568 (1968)

\refno {29.} J. Schwinger, Phys. Lett. {\bf 24B} 473 (1967)

\refno {30.} P. Chang and F. G\"ursey, Phys. Rev. {\bf 164} 1752 (1967)

\refno {31.} R. Jacob and M. D. Scadron, Phys. Rev. D {\bf 25} 3073 (1982)

\refno {32.} J. Gasser and H. Leutwyler, Phys. Lett. {\bf 125B} 325 (1983)

\refno {33.} A. N. Ivanov and N. I. Troitskaya, Yad. Fiz. {\bf 43} 405 (1986)
[Sov. J. Nucl. Phys. {\bf 43} 260 (1986)]

\refno {34.} D. Lohse, J. W. Durso, K. Holinde and J. Speth, Nucl. Phys.
{\bf A516} 513 (1990)

\refno {35} M. C. Ruivo, C. A. de Sousa, B. Hiller and A. H. Blin,
Nucl. Phys. {\bf A575} 460 (1994)

\refno {36} V. Bernard, U.-G. Mei{\ss}ner, A. H. Blin and B. Hiller,
Phys. Lett. {\bf B253} 443 (1991)

\refno {37.} Y. Kuramashi, M. Fukugita, H. Mino, M. Okawa and A. Ukawa,
Phys. Rev. Lett. {\bf 71} 2387 (1993)

\refno {38.} C. D. Roberts, R. T. Cahill, M. E. Sevior and N. Iannella,
Phys. Rev. D {\bf 49} 125 (1994)

\refno {39.} J. R. Uretsky and T. R. Palfrey, Phys. Rev. {\bf 121} 1798 (1961)

\refno {40.} V. Bernard, N. Kaiser and Ulf-G. Mei{\ss}ner, preprint
hep-ph/9404236 (1994)

\refno {41.} A. Zylberstejn, Ph.D. thesis, University of Paris, Orsay, 1972
(unpublished)

\refno {42.} E. W. Beier {\it et al.}, Phys. Rev. Lett. {\bf 29} 511 (1972);
{\it ibid.} {\bf 30} 399 (1973)

\endref
\byebye

\magnification=\magstep0
\font \authfont               = cmr10 scaled\magstep4
\font \fivesans               = cmss10 at 5pt
\font \headfont               = cmbx12 scaled\magstep4
\font \markfont               = cmr10 scaled\magstep1
\font \ninebf                 = cmbx9
\font \ninei                  = cmmi9
\font \nineit                 = cmti9
\font \ninerm                 = cmr9
\font \ninesans               = cmss10 at 9pt
\font \ninesl                 = cmsl9
\font \ninesy                 = cmsy9
\font \ninett                 = cmtt9
\font \sevensans              = cmss10 at 7pt
\font \sixbf                  = cmbx6
\font \sixi                   = cmmi6
\font \sixrm                  = cmr6
\font \sixsans                = cmss10 at 6pt
\font \sixsy                  = cmsy6
\font \smallescriptfont       = cmr5 at 7pt
\font \smallescriptscriptfont = cmr5
\font \smalletextfont         = cmr5 at 10pt
\font \subhfont               = cmr10 scaled\magstep4
\font \tafonts                = cmbx7  scaled\magstep2
\font \tafontss               = cmbx5  scaled\magstep2
\font \tafontt                = cmbx10 scaled\magstep2
\font \tams                   = cmmib10
\font \tamss                  = cmmib10 scaled 700
\font \tamt                   = cmmib10 scaled\magstep2
\font \tass                   = cmsy7  scaled\magstep2
\font \tasss                  = cmsy5  scaled\magstep2
\font \tast                   = cmsy10 scaled\magstep2
\font \tasys                  = cmex10 scaled\magstep1
\font \tasyt                  = cmex10 scaled\magstep2
\font \tbfonts                = cmbx7  scaled\magstep1
\font \tbfontss               = cmbx5  scaled\magstep1
\font \tbfontt                = cmbx10 scaled\magstep1
\font \tbms                   = cmmib10 scaled 833
\font \tbmss                  = cmmib10 scaled 600
\font \tbmt                   = cmmib10 scaled\magstep1
\font \tbss                   = cmsy7  scaled\magstep1
\font \tbsss                  = cmsy5  scaled\magstep1
\font \tbst                   = cmsy10 scaled\magstep1
\font \tenbfne                = cmb10
\font \tensans                = cmss10
\font \tpfonts                = cmbx7  scaled\magstep3
\font \tpfontss               = cmbx5  scaled\magstep3
\font \tpfontt                = cmbx10 scaled\magstep3
\font \tpmt                   = cmmib10 scaled\magstep3
\font \tpss                   = cmsy7  scaled\magstep3
\font \tpsss                  = cmsy5  scaled\magstep3
\font \tpst                   = cmsy10 scaled\magstep3
\font \tpsyt                  = cmex10 scaled\magstep3
\vsize=19.3cm
\hsize=12.2cm
\hfuzz=2pt
\tolerance=500
\abovedisplayskip=3 mm plus6pt minus 4pt
\belowdisplayskip=3 mm plus6pt minus 4pt
\abovedisplayshortskip=0mm plus6pt minus 2pt
\belowdisplayshortskip=2 mm plus4pt minus 4pt
\predisplaypenalty=0
\clubpenalty=10000
\widowpenalty=10000
\frenchspacing
\newdimen\oldparindent\oldparindent=1.5em
\parindent=1.5em
\skewchar\ninei='177 \skewchar\sixi='177
\skewchar\ninesy='60 \skewchar\sixsy='60
\hyphenchar\ninett=-1
\def\newline{\hfil\break}%
\catcode`@=11
\def\folio{\ifnum\pageno<\z@
\uppercase\expandafter{\romannumeral-\pageno}%
\else\number\pageno \fi}
\catcode`@=12 
  \mathchardef\Gamma="0100
  \mathchardef\Delta="0101
  \mathchardef\Theta="0102
  \mathchardef\Lambda="0103
  \mathchardef\Xi="0104
  \mathchardef\Pi="0105
  \mathchardef\Sigma="0106
  \mathchardef\Upsilon="0107
  \mathchardef\Phi="0108
  \mathchardef\Psi="0109
  \mathchardef\Omega="010A
  \mathchardef\bfGamma="0\the\bffam 00
  \mathchardef\bfDelta="0\the\bffam 01
  \mathchardef\bfTheta="0\the\bffam 02
  \mathchardef\bfLambda="0\the\bffam 03
  \mathchardef\bfXi="0\the\bffam 04
  \mathchardef\bfPi="0\the\bffam 05
  \mathchardef\bfSigma="0\the\bffam 06
  \mathchardef\bfUpsilon="0\the\bffam 07
  \mathchardef\bfPhi="0\the\bffam 08
  \mathchardef\bfPsi="0\the\bffam 09
  \mathchardef\bfOmega="0\the\bffam 0A

\def\sq{\hbox{\rlap{$\sqcap$}$\sqcup$}}

\def\utw{\smash{\rlap{\lower5pt\hbox{$\sim$}}}}
\def\udtw{\smash{\rlap{\lower6pt\hbox{$\approx$}}}}

\def\diameter{{\ifmmode\mathchoice
{\ooalign{\hfil\hbox{$\displaystyle/$}\hfil\crcr
{\hbox{$\displaystyle\mathchar"20D$}}}}
{\ooalign{\hfil\hbox{$\textstyle/$}\hfil\crcr
{\hbox{$\textstyle\mathchar"20D$}}}}
{\ooalign{\hfil\hbox{$\scriptstyle/$}\hfil\crcr
{\hbox{$\scriptstyle\mathchar"20D$}}}}
{\ooalign{\hfil\hbox{$\scriptscriptstyle/$}\hfil\crcr
{\hbox{$\scriptscriptstyle\mathchar"20D$}}}}
\else{\ooalign{\hfil/\hfil\crcr\mathhexbox20D}}%
\fi}}


\def\bbbc{{\mathchoice {\setbox0=\hbox{$\displaystyle\rm C$}\hbox{\hbox
to0pt{\kern0.4\wd0\vrule height0.9\ht0\hss}\box0}}
{\setbox0=\hbox{$\textstyle\rm C$}\hbox{\hbox
to0pt{\kern0.4\wd0\vrule height0.9\ht0\hss}\box0}}
{\setbox0=\hbox{$\scriptstyle\rm C$}\hbox{\hbox
to0pt{\kern0.4\wd0\vrule height0.9\ht0\hss}\box0}}
{\setbox0=\hbox{$\scriptscriptstyle\rm C$}\hbox{\hbox
to0pt{\kern0.4\wd0\vrule height0.9\ht0\hss}\box0}}}}
\def\bbbe{{\mathchoice {\setbox0=\hbox{\smalletextfont e}\hbox{\raise
0.1\ht0\hbox to0pt{\kern0.4\wd0\vrule width0.3pt height0.7\ht0\hss}\box0}}
{\setbox0=\hbox{\smalletextfont e}\hbox{\raise
0.1\ht0\hbox to0pt{\kern0.4\wd0\vrule width0.3pt height0.7\ht0\hss}\box0}}
{\setbox0=\hbox{\smallescriptfont e}\hbox{\raise
0.1\ht0\hbox to0pt{\kern0.5\wd0\vrule width0.2pt height0.7\ht0\hss}\box0}}
{\setbox0=\hbox{\smallescriptscriptfont e}\hbox{\raise
0.1\ht0\hbox to0pt{\kern0.4\wd0\vrule width0.2pt height0.7\ht0\hss}\box0}}}}
\def\bbbq{{\mathchoice {\setbox0=\hbox{$\displaystyle\rm Q$}\hbox{\raise
0.15\ht0\hbox to0pt{\kern0.4\wd0\vrule height0.8\ht0\hss}\box0}}
{\setbox0=\hbox{$\textstyle\rm Q$}\hbox{\raise
0.15\ht0\hbox to0pt{\kern0.4\wd0\vrule height0.8\ht0\hss}\box0}}
{\setbox0=\hbox{$\scriptstyle\rm Q$}\hbox{\raise
0.15\ht0\hbox to0pt{\kern0.4\wd0\vrule height0.7\ht0\hss}\box0}}
{\setbox0=\hbox{$\scriptscriptstyle\rm Q$}\hbox{\raise
0.15\ht0\hbox to0pt{\kern0.4\wd0\vrule height0.7\ht0\hss}\box0}}}}
\def\bbbt{{\mathchoice {\setbox0=\hbox{$\displaystyle\rm
T$}\hbox{\hbox to0pt{\kern0.3\wd0\vrule height0.9\ht0\hss}\box0}}
{\setbox0=\hbox{$\textstyle\rm T$}\hbox{\hbox
to0pt{\kern0.3\wd0\vrule height0.9\ht0\hss}\box0}}
{\setbox0=\hbox{$\scriptstyle\rm T$}\hbox{\hbox
to0pt{\kern0.3\wd0\vrule height0.9\ht0\hss}\box0}}
{\setbox0=\hbox{$\scriptscriptstyle\rm T$}\hbox{\hbox
to0pt{\kern0.3\wd0\vrule height0.9\ht0\hss}\box0}}}}
\def\bbbs{{\mathchoice
{\setbox0=\hbox{$\displaystyle     \rm S$}\hbox{\raise0.5\ht0\hbox
to0pt{\kern0.35\wd0\vrule height0.45\ht0\hss}\hbox
to0pt{\kern0.55\wd0\vrule height0.5\ht0\hss}\box0}}
{\setbox0=\hbox{$\textstyle        \rm S$}\hbox{\raise0.5\ht0\hbox
to0pt{\kern0.35\wd0\vrule height0.45\ht0\hss}\hbox
to0pt{\kern0.55\wd0\vrule height0.5\ht0\hss}\box0}}
{\setbox0=\hbox{$\scriptstyle      \rm S$}\hbox{\raise0.5\ht0\hbox
to0pt{\kern0.35\wd0\vrule height0.45\ht0\hss}\raise0.05\ht0\hbox
to0pt{\kern0.5\wd0\vrule height0.45\ht0\hss}\box0}}
{\setbox0=\hbox{$\scriptscriptstyle\rm S$}\hbox{\raise0.5\ht0\hbox
to0pt{\kern0.4\wd0\vrule height0.45\ht0\hss}\raise0.05\ht0\hbox
to0pt{\kern0.55\wd0\vrule height0.45\ht0\hss}\box0}}}}
\def\bbbz{{\mathchoice {\hbox{$\sans\textstyle Z\kern-0.4em Z$}}
{\hbox{$\sans\textstyle Z\kern-0.4em Z$}}
{\hbox{$\sans\scriptstyle Z\kern-0.3em Z$}}
{\hbox{$\sans\scriptscriptstyle Z\kern-0.2em Z$}}}}
\def\qed{\ifmmode\sq\else{\unskip\nobreak\hfil
\penalty50\hskip1em\null\nobreak\hfil\sq
\parfillskip=0pt\finalhyphendemerits=0\endgraf}\fi}
\newfam\sansfam
\textfont\sansfam=\tensans\scriptfont\sansfam=\sevensans
\scriptscriptfont\sansfam=\fivesans
\def\sans{\fam\sansfam\tensans}
\def\stackfigbox{\if
Y\FIG\global\setbox\figbox=\vbox{\unvbox\figbox\box1}%
\else\global\setbox\figbox=\vbox{\box1}\global\let\FIG=Y\fi}
\def\placefigure{\dimen0=\ht1\advance\dimen0by\dp1
\advance\dimen0by5\baselineskip
\advance\dimen0by0.33333 cm
\ifdim\dimen0>\vsize\pageinsert\box1\vfill\endinsert
\else
\if Y\FIG\stackfigbox\else
\dimen0=\pagetotal\ifdim\dimen0<\pagegoal
\advance\dimen0by\ht1\advance\dimen0by\dp1\advance\dimen0by1.16666cm
\ifdim\dimen0>\pagegoal\stackfigbox
\else\box1\vskip3.33333 mm\fi
\else\box1\vskip3.33333 mm\fi\fi\fi}
%
\def\begfig#1cm#2\endfig{\par
\setbox1=\vbox{\dimen0=#1true cm\advance\dimen0
by0.83333 cm\kern\dimen0#2}\placefigure}
\def\begdoublefig#1cm #2 #3 \enddoublefig{\begfig#1cm%
\vskip-.8333\baselineskip\line{\vtop{\hsize=0.46\hsize#2}\hfill
\vtop{\hsize=0.46\hsize#3}}\endfig}
\def\begfigsidebottom#1cm#2cm#3\endfigsidebottom{\dimen0=#2true cm
\ifdim\dimen0<0.4\hsize\message{Room for legend to narrow;
begfigsidebottom changed to begfig}\begfig#1cm#3\endfig\else
\par\def\figure##1##2{\vbox{\noindent\petit{\bf
Fig.\ts##1\unskip.\ }\ignorespaces ##2\par}}%
\dimen0=\hsize\advance\dimen0 by-.66666 cm\advance\dimen0 by-#2true cm
\setbox1=\vbox{\hbox{\hbox to\dimen0{\vrule height#1true cm\hrulefill}%
\kern.66666 cm\vbox{\hsize=#2true cm#3}}}\placefigure\fi}
\def\begfigsidetop#1cm#2cm#3\endfigsidetop{\dimen0=#2true cm
\ifdim\dimen0<0.4\hsize\message{Room for legend to narrow; begfigsidetop
changed to begfig}\begfig#1cm#3\endfig\else
\par\def\figure##1##2{\vbox{\noindent\petit{\bf
Fig.\ts##1\unskip.\ }\ignorespaces ##2\par}}%
\dimen0=\hsize\advance\dimen0 by-.66666 cm\advance\dimen0 by-#2true cm
\setbox1=\vbox{\hbox{\hbox to\dimen0{\vrule height#1true cm\hrulefill}%
\kern.66666 cm\vbox to#1true cm{\hsize=#2true cm#3\vfill
}}}\placefigure\fi}
\def\figure#1#2{\vskip0.83333 cm\setbox0=\vbox{\noindent\petit{\bf
Fig.\ts#1\unskip.\ }\ignorespaces #2\smallskip
\count255=0\global\advance\count255by\prevgraf}%
\ifnum\count255>1\box0\else
\centerline{\petit{\bf Fig.\ts#1\unskip.\
}\ignorespaces#2}\smallskip\fi}
\def\tabcap#1#2{\smallskip\vbox{\noindent\petit{\bf Table\ts#1\unskip.\
}\ignorespaces #2\medskip}}
\def\begtab#1cm#2\endtab{\par
   \ifvoid\topins\midinsert\medskip\vbox{#2\kern#1true cm}\endinsert
   \else\topinsert\vbox{#2\kern#1true cm}\endinsert\fi}
\def\begpet{\vskip6pt\bgroup\petit}
\def\endpet{\vskip6pt\egroup}
\newcount\frpages
\newcount\frpagegoal
\def\freepage#1{\global\frpagegoal=#1\relax\global\frpages=0\relax
\loop\global\advance\frpages by 1\relax
\message{Doing freepage \the\frpages\space of
\the\frpagegoal}\null\vfill\eject
\ifnum\frpagegoal>\frpages\repeat}
\newdimen\refindent
\def\begrefchapter#1{\titlea{}{\ignorespaces#1}%
\bgroup\petit
\setbox0=\hbox{1000.\enspace}\refindent=\wd0}
\def\ref{\goodbreak
\hangindent\oldparindent\hangafter=1
\noindent\ignorespaces}
\def\refno#1{\goodbreak
\hangindent\refindent\hangafter=1
\noindent\hbox to\refindent{#1\hss}\ignorespaces}
\def\endref{\goodbreak\endpet}
\def\vec#1{{\textfont1=\tams\scriptfont1=\tamss
\textfont0=\tenbf\scriptfont0=\sevenbf
\mathchoice{\hbox{$\displaystyle#1$}}{\hbox{$\textstyle#1$}}
{\hbox{$\scriptstyle#1$}}{\hbox{$\scriptscriptstyle#1$}}}}
\def\petit{\def\rm{\fam0\ninerm}%
\textfont0=\ninerm \scriptfont0=\sixrm \scriptscriptfont0=\fiverm
 \textfont1=\ninei \scriptfont1=\sixi \scriptscriptfont1=\fivei
 \textfont2=\ninesy \scriptfont2=\sixsy \scriptscriptfont2=\fivesy
 \def\it{\fam\itfam\nineit}%
 \textfont\itfam=\nineit
 \def\sl{\fam\slfam\ninesl}%
 \textfont\slfam=\ninesl
 \def\bf{\fam\bffam\ninebf}%
 \textfont\bffam=\ninebf \scriptfont\bffam=\sixbf
 \scriptscriptfont\bffam=\fivebf
 \def\sans{\fam\sansfam\ninesans}%
 \textfont\sansfam=\ninesans \scriptfont\sansfam=\sixsans
 \scriptscriptfont\sansfam=\fivesans
 \def\tt{\fam\ttfam\ninett}%
 \textfont\ttfam=\ninett
 \normalbaselineskip=11pt
 \setbox\strutbox=\hbox{\vrule height7pt depth2pt width0pt}%
 \normalbaselines\rm
\def\vec##1{{\textfont1=\tbms\scriptfont1=\tbmss
\textfont0=\ninebf\scriptfont0=\sixbf
\mathchoice{\hbox{$\displaystyle##1$}}{\hbox{$\textstyle##1$}}
{\hbox{$\scriptstyle##1$}}{\hbox{$\scriptscriptstyle##1$}}}}}
\nopagenumbers
%
\let\header=Y
\let\FIG=N
\newbox\figbox
\output={\if N\header\headline={\hfil}\fi\plainoutput\global\let\header=Y
\if Y\FIG\topinsert\unvbox\figbox\endinsert\global\let\FIG=N\fi}
\let\lasttitle=N
\def\bookauthor#1{\vfill\eject
     \bgroup
     \baselineskip=22pt
     \lineskip=0pt
     \pretolerance=10000
     \authfont
     \rightskip 0pt plus 6em
     \centerpar{#1}\vskip1.66666 cm\egroup}
\def\bookhead#1#2{\bgroup
     \baselineskip=36pt
     \lineskip=0pt
     \pretolerance=10000
     \headfont
     \rightskip 0pt plus 6em
     \centerpar{#1}\vskip0.83333 cm
     \baselineskip=22pt
     \subhfont\centerpar{#2}\vfill
     \parindent=0pt
     \baselineskip=16pt
     \leftskip=1.83333cm
     \markfont Springer-Verlag\newline
     Berlin Heidelberg New York\newline
     London Paris Tokyo Singapore\bigskip\bigskip
     [{\it This is page III of your manuscript and will be reset by
     Springer.}]
     \egroup\let\header=N\eject}
\def\centerpar#1{{\parfillskip=0pt
\rightskip=0pt plus 1fil
\leftskip=0pt plus 1fil
\advance\leftskip by\oldparindent
\advance\rightskip by\oldparindent
\def\newline{\break}%
\noindent\ignorespaces#1\par}}
\def\part#1#2{\vfill\supereject\let\header=N
\centerline{\subhfont#1}%
\vskip75pt
     \bgroup
\textfont0=\tpfontt \scriptfont0=\tpfonts \scriptscriptfont0=\tpfontss
\textfont1=\tpmt \scriptfont1=\tbmt \scriptscriptfont1=\tams
\textfont2=\tpst \scriptfont2=\tpss \scriptscriptfont2=\tpsss
\textfont3=\tpsyt \scriptfont3=\tasys \scriptscriptfont3=\tenex
     \baselineskip=20pt
     \lineskip=0pt
     \pretolerance=10000
     \tpfontt
     \centerpar{#2}
     \vfill\eject\egroup\ignorespaces}
\newtoks\AUTHOR
\newtoks\HEAD
\catcode`\@=\active
\def\author#1{\bgroup
\baselineskip=22pt
\lineskip=0pt
\pretolerance=10000
\markfont
\centerpar{#1}\bigskip\egroup
{\def@##1{}%
\setbox0=\hbox{\petit\kern2.08333 cc\ignorespaces#1\unskip}%
\ifdim\wd0>\hsize
\message{The names of the authors exceed the headline, please use a }%
\message{short form with AUTHORRUNNING}\gdef\leftheadline{%
\hbox to2.08333 cc{\folio\hfil}AUTHORS suppressed due to excessive
length\hfil}%
\global\AUTHOR={AUTHORS were to long}\else
\xdef\leftheadline{\hbox to2.08333
cc{\noexpand\folio\hfil}\ignorespaces#1\hfill}%
\global\AUTHOR={\def@##1{}\ignorespaces#1\unskip}\fi
}\let\INS=E}
\def\address#1{\bgroup
\centerpar{#1}\bigskip\egroup
\catcode`\@=12
\vskip2cm\noindent\ignorespaces}
\let\INS=N%
\def@#1{\if N\INS\unskip\ $^{#1}$\else\if
E\INS\noindent$^{#1}$\let\INS=Y\ignorespaces
\else\par
\noindent$^{#1}$\ignorespaces\fi\fi}%
\catcode`\@=12
\headline={\petit\def\newline{ }\def\fonote#1{}\ifodd\pageno
\rightheadline\else\leftheadline\fi}
\def\rightheadline{\hfil Missing CONTRIBUTION
title\hbox to2.08333 cc{\hfil\folio}}
\def\leftheadline{\hbox to2.08333 cc{\folio\hfil}Missing name(s) of the
author(s)\hfil}
\nopagenumbers
\let\header=Y

\let\lasttitle=N
 \def\contribution#1{\vfill\supereject
 \ifodd\pageno\else\null\vfill\supereject\fi
 \let\header=N\bgroup
 \textfont0=\tafontt \scriptfont0=\tafonts \scriptscriptfont0=\tafontss
 \textfont1=\tamt \scriptfont1=\tams \scriptscriptfont1=\tams
 \textfont2=\tast \scriptfont2=\tass \scriptscriptfont2=\tasss
 \par\baselineskip=16pt
     \lineskip=16pt
     \tafontt
     \raggedright
     \pretolerance=10000
     \noindent
     \centerpar{\ignorespaces#1}%
     \vskip12pt\egroup
     \nobreak
     \parindent=0pt
     \everypar={\global\parindent=1.5em
     \global\let\lasttitle=N\global\everypar={}}%
     \global\let\lasttitle=A%
     \setbox0=\hbox{\petit\def\newline{ }\def\fonote##1{}\kern2.08333
     cc\ignorespaces#1}\ifdim\wd0>\hsize
     \message{Your CONTRIBUTIONtitle exceeds the headline,
please use a short form
with CONTRIBUTIONRUNNING}\gdef\rightheadline{\hfil CONTRIBUTION title
suppressed due to excessive length\hbox to2.08333 cc{\hfil\folio}}%
\global\HEAD={HEAD was to long}\else
\gdef\rightheadline{\hfill\ignorespaces#1\unskip\hbox to2.08333
cc{\hfil\folio}}\global\HEAD={\ignorespaces#1\unskip}\fi
\catcode`\@=\active
     \ignorespaces}
 \def\contributionnext#1{\vfill\supereject
 \let\header=N\bgroup
 \textfont0=\tafontt \scriptfont0=\tafonts \scriptscriptfont0=\tafontss
 \textfont1=\tamt \scriptfont1=\tams \scriptscriptfont1=\tams
 \textfont2=\tast \scriptfont2=\tass \scriptscriptfont2=\tasss
 \par\baselineskip=16pt
     \lineskip=16pt
     \tafontt
     \raggedright
     \pretolerance=10000
     \noindent
     \centerpar{\ignorespaces#1}%
     \vskip12pt\egroup
     \nobreak
     \parindent=0pt
     \everypar={\global\parindent=1.5em
     \global\let\lasttitle=N\global\everypar={}}%
     \global\let\lasttitle=A%
     \setbox0=\hbox{\petit\def\newline{ }\def\fonote##1{}\kern2.08333
     cc\ignorespaces#1}\ifdim\wd0>\hsize
     \message{Your CONTRIBUTIONtitle exceeds the headline,
please use a short form
with CONTRIBUTIONRUNNING}\gdef\rightheadline{\hfil CONTRIBUTION title
suppressed due to excessive length\hbox to2.08333 cc{\hfil\folio}}%
\global\HEAD={HEAD was to long}\else
\gdef\rightheadline{\hfill\ignorespaces#1\unskip\hbox to2.08333
cc{\hfil\folio}}\global\HEAD={\ignorespaces#1\unskip}\fi
\catcode`\@=\active
     \ignorespaces}
\def\motto#1#2{\bgroup\petit\leftskip=5.41666cm\noindent\ignorespaces#1
\if!#2!\else\medskip\noindent\it\ignorespaces#2\fi\bigskip\egroup
\let\lasttitle=M
\parindent=0pt
\everypar={\global\parindent=\oldparindent
\global\let\lasttitle=N\global\everypar={}}%
\global\let\lasttitle=M%
\ignorespaces}
\def\abstract#1{\bgroup\petit\noindent
{\bf Abstract: }\ignorespaces#1\vskip28pt\egroup
\let\lasttitle=N
\parindent=0pt
\everypar={\global\parindent=\oldparindent
\global\let\lasttitle=N\global\everypar={}}%
\ignorespaces}
\def\titlea#1#2{\if N\lasttitle\else\vskip-28pt
     \fi
     \vskip18pt plus 4pt minus4pt
     \bgroup
\textfont0=\tbfontt \scriptfont0=\tbfonts \scriptscriptfont0=\tbfontss
\textfont1=\tbmt \scriptfont1=\tbms \scriptscriptfont1=\tbmss
\textfont2=\tbst \scriptfont2=\tbss \scriptscriptfont2=\tbsss
\textfont3=\tasys \scriptfont3=\tenex \scriptscriptfont3=\tenex
     \baselineskip=16pt
     \lineskip=0pt
     \pretolerance=10000
     \noindent
     \tbfontt
     \rightskip 0pt plus 6em
     \setbox0=\vbox{\vskip23pt\def\fonote##1{}%
     \noindent
     \if!#1!\ignorespaces#2
     \else\setbox0=\hbox{\ignorespaces#1\unskip\ }\hangindent=\wd0
     \hangafter=1\box0\ignorespaces#2\fi
     \vskip18pt}%
     \dimen0=\pagetotal\advance\dimen0 by-\pageshrink
     \ifdim\dimen0<\pagegoal
     \dimen0=\ht0\advance\dimen0 by\dp0\advance\dimen0 by
     3\normalbaselineskip
     \advance\dimen0 by\pagetotal
     \ifdim\dimen0>\pagegoal\eject\fi\fi
     \noindent
     \if!#1!\ignorespaces#2
     \else\setbox0=\hbox{\ignorespaces#1\unskip\ }\hangindent=\wd0
     \hangafter=1\box0\ignorespaces#2\fi
     \vskip18pt plus4pt minus4pt\egroup
     \nobreak
     \parindent=0pt
     \everypar={\global\parindent=\oldparindent
     \global\let\lasttitle=N\global\everypar={}}%
     \global\let\lasttitle=A%
     \ignorespaces}
 \def\titleb#1#2{\if N\lasttitle\else\vskip-28pt
     \fi
     \vskip18pt plus 4pt minus4pt
     \bgroup
\textfont0=\tenbf \scriptfont0=\sevenbf \scriptscriptfont0=\fivebf
\textfont1=\tams \scriptfont1=\tamss \scriptscriptfont1=\tbmss
     \lineskip=0pt
     \pretolerance=10000
     \noindent
     \bf
     \rightskip 0pt plus 6em
     \setbox0=\vbox{\vskip23pt\def\fonote##1{}%
     \noindent
     \if!#1!\ignorespaces#2
     \else\setbox0=\hbox{\ignorespaces#1\unskip\enspace}\hangindent=\wd0
     \hangafter=1\box0\ignorespaces#2\fi
     \vskip10pt}%
     \dimen0=\pagetotal\advance\dimen0 by-\pageshrink
     \ifdim\dimen0<\pagegoal
     \dimen0=\ht0\advance\dimen0 by\dp0\advance\dimen0 by
     3\normalbaselineskip
     \advance\dimen0 by\pagetotal
     \ifdim\dimen0>\pagegoal\eject\fi\fi
     \noindent
     \if!#1!\ignorespaces#2
     \else\setbox0=\hbox{\ignorespaces#1\unskip\enspace}\hangindent=\wd0
     \hangafter=1\box0\ignorespaces#2\fi
     \vskip8pt plus4pt minus4pt\egroup
     \nobreak
     \parindent=0pt
     \everypar={\global\parindent=\oldparindent
     \global\let\lasttitle=N\global\everypar={}}%
     \global\let\lasttitle=B%
     \ignorespaces}
 \def\titlec#1#2{\if N\lasttitle\else\vskip-23pt
     \fi
     \vskip18pt plus 4pt minus4pt
     \bgroup
\textfont0=\tenbfne \scriptfont0=\sevenbf \scriptscriptfont0=\fivebf
\textfont1=\tams \scriptfont1=\tamss \scriptscriptfont1=\tbmss
     \tenbfne
     \lineskip=0pt
     \pretolerance=10000
     \noindent
     \rightskip 0pt plus 6em
     \setbox0=\vbox{\vskip23pt\def\fonote##1{}%
     \noindent
     \if!#1!\ignorespaces#2
     \else\setbox0=\hbox{\ignorespaces#1\unskip\enspace}\hangindent=\wd0
     \hangafter=1\box0\ignorespaces#2\fi
     \vskip6pt}%
     \dimen0=\pagetotal\advance\dimen0 by-\pageshrink
     \ifdim\dimen0<\pagegoal
     \dimen0=\ht0\advance\dimen0 by\dp0\advance\dimen0 by
     2\normalbaselineskip
     \advance\dimen0 by\pagetotal
     \ifdim\dimen0>\pagegoal\eject\fi\fi
     \noindent
     \if!#1!\ignorespaces#2
     \else\setbox0=\hbox{\ignorespaces#1\unskip\enspace}\hangindent=\wd0
     \hangafter=1\box0\ignorespaces#2\fi
     \vskip6pt plus4pt minus4pt\egroup
     \nobreak
     \parindent=0pt
     \everypar={\global\parindent=\oldparindent
     \global\let\lasttitle=N\global\everypar={}}%
     \global\let\lasttitle=C%
     \ignorespaces}
 \def\titled#1{\if N\lasttitle\else\vskip-\baselineskip
     \fi
     \vskip12pt plus 4pt minus 4pt
     \bgroup
\textfont1=\tams \scriptfont1=\tamss \scriptscriptfont1=\tbmss
     \bf
     \noindent
     \ignorespaces#1\ \ignorespaces\egroup
     \ignorespaces}
\let\ts=\thinspace
\def\footnoterule{\kern-3pt\hrule width 1.66666 cm\kern2.6pt}
\newcount\footcount \footcount=0
\def\advftncnt{\advance\footcount by1\global\footcount=\footcount}
\def\fonote#1{\advftncnt$^{\the\footcount}$\begingroup\petit
\parfillskip=0pt plus 1fil
\def\textindent##1{\hangindent0.5\oldparindent\noindent\hbox
to0.5\oldparindent{##1\hss}\ignorespaces}%
\vfootnote{$^{\the\footcount}$}{#1\vskip-9.69pt}\endgroup}
\def\item#1{\par\noindent
\hangindent6.5 mm\hangafter=0
\llap{#1\enspace}\ignorespaces}

\def\titleao#1{\vfill\supereject
\ifodd\pageno\else\null\vfill\supereject\fi
\let\header=N
     \bgroup
\textfont0=\tafontt \scriptfont0=\tafonts \scriptscriptfont0=\tafontss
\textfont1=\tamt \scriptfont1=\tams \scriptscriptfont1=\tamss
\textfont2=\tast \scriptfont2=\tass \scriptscriptfont2=\tasss
\textfont3=\tasyt \scriptfont3=\tasys \scriptscriptfont3=\tenex
     \baselineskip=18pt
     \lineskip=0pt
     \pretolerance=10000
     \tafontt
     \centerpar{#1}%
     \vskip75pt\egroup
     \nobreak
     \parindent=0pt
     \everypar={\global\parindent=\oldparindent
     \global\let\lasttitle=N\global\everypar={}}%
     \global\let\lasttitle=A%
     \ignorespaces}






\def\leaderfill{\kern0.5em\leaders\hbox to 0.5em{\hss.\hss}\hfill\kern
0.5em}
\newdimen\chapindent
\newdimen\sectindent
\newdimen\subsecindent
\newdimen\thousand
\setbox0=\hbox{\bf 10. }\chapindent=\wd0
\setbox0=\hbox{10.10 }\sectindent=\wd0
\setbox0=\hbox{10.10.1 }\subsecindent=\wd0
\setbox0=\hbox{\enspace 100}\thousand=\wd0
\def\contpart#1#2{\medskip\noindent
\vbox{\kern10pt\leftline{\textfont1=\tams
\scriptfont1=\tamss\scriptscriptfont1=\tbmss\bf
\advance\chapindent by\sectindent
\hbox to\chapindent{\ignorespaces#1\hss}\ignorespaces#2}\kern8pt}%
\let\lasttitle=Y\par}
\def\contcontribution#1#2{\if N\lasttitle\bigskip\fi
\let\lasttitle=N\line{{\textfont1=\tams
\scriptfont1=\tamss\scriptscriptfont1=\tbmss\bf#1}%
\if!#2!\hfill\else\leaderfill\hbox to\thousand{\hss#2}\fi}\par}
\def\conttitlea#1#2#3{\line{\hbox to
\chapindent{\strut\bf#1\hss}{\textfont1=\tams
\scriptfont1=\tamss\scriptscriptfont1=\tbmss\bf#2}%
\if!#3!\hfill\else\leaderfill\hbox to\thousand{\hss#3}\fi}\par}
\def\conttitleb#1#2#3{\line{\kern\chapindent\hbox
to\sectindent{\strut#1\hss}{#2}%
\if!#3!\hfill\else\leaderfill\hbox to\thousand{\hss#3}\fi}\par}
\def\conttitlec#1#2#3{\line{\kern\chapindent\kern\sectindent
\hbox to\subsecindent{\strut#1\hss}{#2}%
\if!#3!\hfill\else\leaderfill\hbox to\thousand{\hss#3}\fi}\par}
\long\def\lemma#1#2{\removelastskip\vskip\baselineskip\noindent{\tenbfne
Lemma\if!#1!\else\ #1\fi\ \ }{\it\ignorespaces#2}\vskip\baselineskip}
\long\def\proposition#1#2{\removelastskip\vskip\baselineskip\noindent{\tenbfne
Proposition\if!#1!\else\ #1\fi\ \ }{\it\ignorespaces#2}\vskip\baselineskip}
\long\def\theorem#1#2{\removelastskip\vskip\baselineskip\noindent{\tenbfne
Theorem\if!#1!\else\ #1\fi\ \ }{\it\ignorespaces#2}\vskip\baselineskip}
\long\def\corollary#1#2{\removelastskip\vskip\baselineskip\noindent{\tenbfne
Corollary\if!#1!\else\ #1\fi\ \ }{\it\ignorespaces#2}\vskip\baselineskip}
\long\def\example#1#2{\removelastskip\vskip\baselineskip\noindent{\tenbfne
Example\if!#1!\else\ #1\fi\ \ }\ignorespaces#2\vskip\baselineskip}
\long\def\exercise#1#2{\removelastskip\vskip\baselineskip\noindent{\tenbfne
Exercise\if!#1!\else\ #1\fi\ \ }\ignorespaces#2\vskip\baselineskip}
\long\def\problem#1#2{\removelastskip\vskip\baselineskip\noindent{\tenbfne
Problem\if!#1!\else\ #1\fi\ \ }\ignorespaces#2\vskip\baselineskip}
\long\def\solution#1#2{\removelastskip\vskip\baselineskip\noindent{\tenbfne
Solution\if!#1!\else\ #1\fi\ \ }\ignorespaces#2\vskip\baselineskip}


\long\def\definition#1#2{\removelastskip\vskip\baselineskip\noindent{\tenbfne
Definition\if!#1!\else\
#1\fi\ \ }\ignorespaces#2\vskip\baselineskip}
\def\frame#1{\bigskip\vbox{\hrule\hbox{\vrule\kern5pt
\vbox{\kern5pt\advance\hsize by-10.8pt
\centerline{\vbox{#1}}\kern5pt}\kern5pt\vrule}\hrule}\bigskip}
\def\frameddisplay#1#2{$$\vcenter{\hrule\hbox{\vrule\kern5pt
\vbox{\kern5pt\hbox{$\displaystyle#1$}%
\kern5pt}\kern5pt\vrule}\hrule}\eqno#2$$}
\def\typeset{\petit\noindent This book was processed by the author using
the \TeX\ macro package from Springer-Verlag.\par}
\outer\def\byebye{\bigskip\bigskip\typeset
\footcount=1\ifx\speciali\undefined\else
\loop\smallskip\noindent special character No\number\footcount:
\csname special\romannumeral\footcount\endcsname
\advance\footcount by 1\global\footcount=\footcount
\ifnum\footcount<11\repeat\fi
\gdef\leftheadline{\hbox to2.08333 cc{\folio\hfil}\ignorespaces
\the\AUTHOR\unskip: \the\HEAD\hfill}\vfill\supereject\end}